\documentclass[printer]{aa} 

\usepackage{graphicx}
\usepackage{url}
\usepackage{txfonts}
\usepackage{natbib}
\bibpunct{(}{)}{;}{a}{}{,} 
\begin{document}

   \title{Study of the molecular and ionized gas in a possible precursor of an ultra-compact \ion{H}{II} region}

\author {M. E. Ortega \inst{1}
\and S. Paron \inst{1,2}
\and E. Giacani \inst{1,2}
\and M. Celis Pe\~na \inst{1}
\and M. Rubio \inst{3}
\and A. Petriella \inst{1}
}

\institute{CONICET - Universidad de Buenos Aires, Instituto de Astronom\'{\i}a y F\'{\i}sica del Espacio (IAFE),
             CP 1428 Buenos Aires, Argentina\\
             \email{mortega@iafe.uba.ar}
\and Universidad de Buenos Aires, Facultad de Arquitectura, Dise\~{n}o y Urbanismo, Departamento de Dise\~{n}o Industrial, CP 1421, Buenos Aires, Argentina
\and Departamento de Astronom\'{\i}a, Universidad de Chile, Casilla 36-D, Santiago, Chile
}

   \date{Received XXX; accepted XXX}

 
  \abstract
   {}  
   {We study the molecular and the ionized gas in a possible precursor of an ultra-compact \ion{H}{II} region to contribute to the understanding of how high-mass stars build-up their masses once they have reached the zero-age main secuence.}
   {We carried out molecular observations towards the position of the Red MSX source G052.9221$-$00.4892, using the Atacama Submillimeter Telescope Experiment (ASTE; Chile) in the $^{12}$CO J=3$-$2, $^{13}$CO J=3$-$2, C$^{18}$O J=3$-$2, and HCO$^+$ J=4$-$3 lines with an angular resolution of about 22$^{\prime\prime}$. We also present radio continuum observations at 6~GHz carried out with the Jansky Very Large Array (JVLA; USA) interferometer with a synthesized beam of $4\farcs 8\times 4\farcs 1$. The molecular data were used to study the distribution and kinematics of the molecular gas, while the radio continuum data were used to characterize the ionized gas in the region. Combining these observations with public infrared data allowed us to inquire about the nature of the source.}
   {The analysis of the molecular observations reveals the presence of a kinetic temperature and H$_2$ column density gradients
across the molecular clump in which the Red MSX source G052.9221$-$00.4892 is embedded, with the hotter and less dense gas in the inner region. The $^{12}$CO J=3$-$2 emission shows evidence of misaligned massive molecular outflows, with the blue lobe in positional coincidence  with a jet-like feature seen at 8~$\mu$m. The radio continuum emission shows a slightly elongated compact radio source, with a flux density of about 0.9~mJy, in positional coincidence with the Red MSX source. The polar-like morphology of this compact radio source perfectly matches the hourglass-like morphology exhibited by the source in the K$_s-$band. Moreover, the axes of symmetry of the radio source and the near-infrared nebula are perfectly aligned. Thus, based on the presence of molecular outflows, the slightly elongated morphology of the compact radio source matching the hourglass-like morphology of the source at the K$_s-$band, and the lack of evidence of collimated jets in the near-infrared spectrum, one interpretation for the nature of the source, is that the Red MSX source G052.9221$-$00.4892 could be transiting a hyper-compact \ion{H}{II} region phase, in which the young central star emits winds and ionizing radiation through the poles. 

By the other hand, according to a comparison between the Br$\gamma$ intensity and the radio flux density at 6~GHz, the source would be in a more evolved evolutionary stage of an optically thin UC \ion{H}{II} region in photoionization equilibrium. If this is the case, from the radio continuum emission, we can conjecture upon the spectral type of its exciting star which would be a B0.5V. 
}
   {}

   \keywords{ISM: clouds -- Stars: formation -- Stars: winds, outflows}

   \maketitle
%

\section{Introduction}

The last few years have seen a rapidly growing observational activity aimed at the characterization of high-mass star forming regions exhibiting a wide range of evolutionary stages that go from the `Hot Cores' \citep{ces94} to the ultra-compact (UC) \ion{H}{II} regions \citep{woo89}. However, the formation of high-mass stars is not well understood yet \citep[e.g.][]{zin07, tan14}, mainly because their earliest stages of evolution have typical timescales of about $10^5$~yrs \citep{tan02}. 
At present, two theoretical scenarios are proposed to explain the formation of these stars: a monolithic collapse of turbulent gas on the scale of massive dense cores \citep{tan02}, which is a scaled-up version of the low mass stars formation picture, and a competitive one where accretion occurs inside the gravitational potential of a cluster-forming massive dense core \citep{bon06}.

The currently accepted evolutionary path of high-mass stars begins inside dense and massive molecular cores. The young stars finish their contraction and reach the zero age main sequence (ZAMS) very rapidly \citep{ket06}. At this point, the star begins to radiate extreme ultraviolet photons which ionize its surroundings, generating a hyper-compact (HC) \ion{H}{II} region \citep{kur05}. For the accretion to prevail against the radiation pressure, an accretion rate of several orders of magnitude above the typical values related to low-mass star formation is required \citep{gar99}. However, what seems to be crucial in this issue is a non-spherical accretion, with the accreting material reaching the young star by flowing inwards, mainly through the equatorial plane \citep[e.g.][]{kui11}. \citet{ket07}, based on analytical calculations, characterized different evolutionary stages of the HC \ion{H}{II} region phase. In a first stage, arises a small quasi-spherical \ion{H}{II} region gravitationally trapped due to the accretion flow. Then, in a second stage, the ionization increases and the \ion{H}{II} region transitions to a bipolar morphology in which appears outward flows of ionized gas through the poles direction.  Thus, the study of the HC \ion{H}{II} regions is crucial to understand, for example, how O-type stars acquire about the half of the final stellar mass after the star begins to produce ionizing radiation \citep{zha14}. However, given the short lifetime of this stage and the biases toward the detection of UC \ion{H}{II} regions of most radio continuum surveys, the number of HC \ion{H}{II} regions cataloged is still scarce.

In this work, we report on the study of the molecular and ionized gas associated with the Red MSX source G052.9221$-$00.4892 (hereafter MSXG52). MSXG52 has been cataloged as an \ion{H}{II} region in the Red MSX Source Survey \citep{lum13}, based on its associated radio continuum emission \citep[CORNISH Survey;][]{hoa12}. MSXG52 is embedded in a pillar-like structure located onto the border of the infrared dust bubble MWP1G052845$-$005363 (hereafter bubble G52) first cataloged by \citet{sim12}. This irregular bubble of about 10~arcmin in size and centered at l=52.845; b=$-$0.536,  is the infrared counterpart of the \ion{H}{II} region G052.9$-$00.6, which has a radio recombination line at  43.5~kms$^{-1}$ \citep{loc89}. Using a flat rotation model for our Galaxy (with R = 7.6 $\pm$ 0.3~kpc and $\theta$ =214 $\pm$ 7~kms$^{-1}$) this velocity corresponds to the near and far distances of about 5.1~kpc (tangent point) and 7.2~kpc, respectively. \citet{and09} resolved the kinematic distance ambiguity in favor of the far distance based on the HI self-absorption method. Thus, in what follows we assume 7.2~kpc as the distance to the bubble G52 and  MSXG52.

\section{Presentation of bubble G52, the pillar, and MSXG52}
\label{presentation}

   \begin{figure}
   \centering
   \includegraphics[width=9cm]{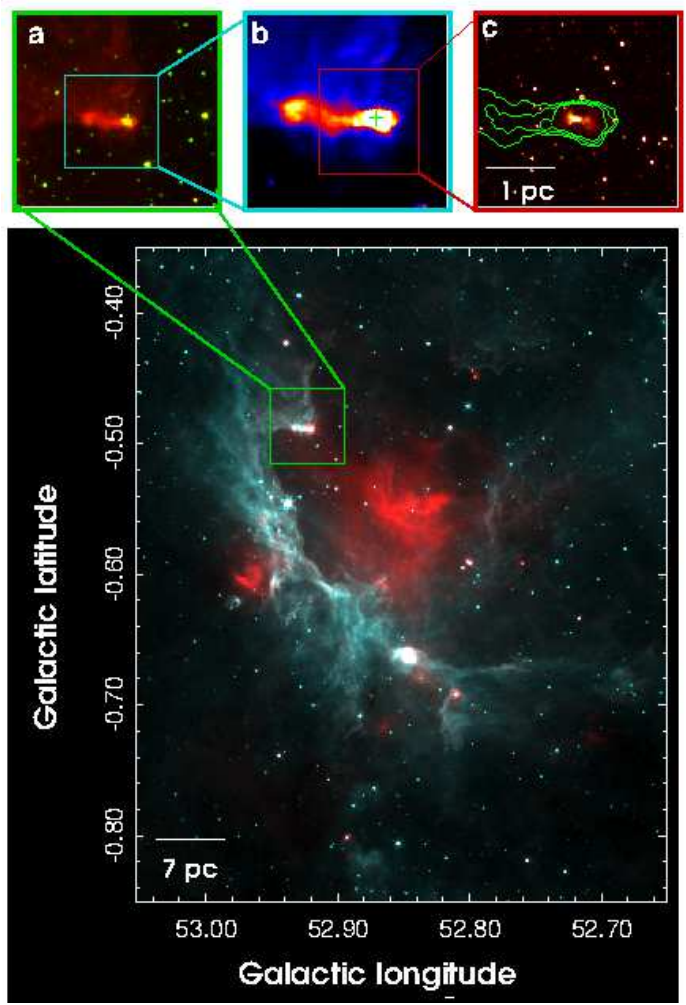}
   \caption{{\it Spitzer} two-color image (24~$\mu$m in red and 8~$\mu$m in cyan)
    of bubble G52 (infrared counterpart of the \ion{H}{II} region
     G052.9$-$00.6). (a) Zoom-up view of the pillar-like
     feature with  4.5~$\mu$m in green, and 8~$\mu$m in red. Both scales go between 20 and 150~MJy~beam$^{-1}$. (b) Zoom-up view of a jet-like structure seen at 8~$\mu$m. The cross represents the position of MSXG52. (c) Zoom-up view of the proto-star candidate seen at $K_s-$band extracted from the UKIDSS survey. The green contours represent the 8~ $\mu$m emission levels at 80, 100, and 140~MJy/beam. }
              \label{intro}
    \end{figure}
    
Figure \ref{intro} shows a two-color composite {\it Spitzer} image (24~$\mu$m in red and 8~$\mu$m in cyan) of bubble G52.  The 24~$\mu$m emission \citep[MIPSGAL;][]{car09}, which might be associated with very small grains, is confined to the interior of the bubble. The 8~$\mu$m emission \citep[GLIMPSE;][]{ben03}, which traces the photo-dissociation regions (PDRs), exhibits a semi-shell like morphology opened towards lower Galactic longitudes. The most interesting feature at this band is a structure with a conspicuous pillar morphology  (green box in the figure), which seems to have been sculpted by the action of the \ion{H}{II} region. Figure \ref{intro}-(a) shows a zoom-up view of this structure in a {\it Spitzer} two-color image where the 4.5 and 8~$\mu$m emissions are represented in green and red, respectively. Towards the head of the pillar, it can be appreciated a slightly elongated bright bulk of emission (seen in yellow) that corresponds to the position of  MSXG52. In connection with this source, appears a curved filament that resembles a typical head-tail jet morphology pointing towards higher Galactic longitudes.  Besides, the Figure \ref{intro}-(a) shows the remarkable prominence of the 8~$\mu$m emission with respect to the 4.5~$\mu$m emission (which is most probably attributed to shocked H$_2$) in the jet-like feature, which suggests the presence of hydrocarbons heated by FUV photons from the central source that would be illuminating  cavity walls \citep[e.g.][]{qiu08, vanden00}. 
Besides, reflection nebulae dominated by the emission from FUV-heated hydrocarbons in the 8~$\mu$m band may imply the presence of young B stars \citep{qiu08}. 
Figure \ref{intro}-(b) shows a zoom-up view of the jet-like feature and the central source seen at 8~$\mu$m. The bulk of the emission exhibits a slightly elongated morphology centered at the location of MSXG52 (green cross). Towards higher Galactic longitudes it extends the curved filament which connects the bulk of the emission with a bow-shock-like feature which crowns the structure.  Figure \ref{intro}-(c) shows a zoom-up view of the central source seen at $K_s-$band extracted of the UKIRT Infrared Deep Sky Survey \citep[UKIDSS;][]{war07}. The green contours represent the 8~$\mu$m emission levels at 80, 100, and 140~MJy~beam$^{-1}$. The nebulosity associated with the UKIDSS source UGPS J193054.62$+$172842.0 \citep{luc08} exhibits a hourglass-like morphology with its symmetry axis perfectly aligned with the jet-like feature seen at 8~$\mu$m. This bipolar cone-like shape morphology suggests the presence of cavities cleared in the circumstellar material. These kind of cavities can be originated by the action of a wide-angle wind arising from a proto-star at its latest stages of evolution or a young star that has recently reached the ZAMS \citep{wei06}, or by a precessing jet that cleared the circumstellar medium \citep{kra06}. The cone-nebulosity that points towards the jet-like feature (left-cone hereafter) appears brighter and  more collimated than the other one (right-cone hereafter), which probably, given its location, is evolving towards a less dense medium. This would explain the absence of a jet-like feature at 8~$\mu$m towards lower Galactic longitudes.  

\begin{figure*}
   \centering
   \includegraphics[width=13cm]{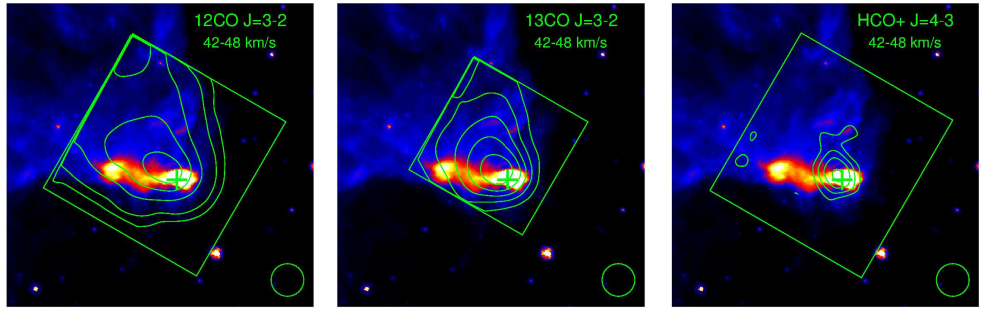}
   \caption{Averaged velocity maps of $^{12}$CO J=3$-$2, $^{13}$CO J=3$-$2, and HCO$^+$ J=4$-$3 emissions between 42 and 48~kms$^{-1}$. The boxes represent the region mapped with the ASTE telescope. The cross indicates the position of the MSXG52. Contours levels are at 1, 2, 3, 5, and 7~K for $^{12}$CO J=3$-$2, at 1, 1.5, 2, 2.5, 3, 3.5~K for $^{13}$CO J=3$-$2, and at 0.4, 0.6, 0.9, and 1.2~K for HCO$^+$ J=4$-$3.  Colour scale goes between 20 and 150~MJy~beam$^{-1}$}.
     \label{integrados}
    \end{figure*}

Motivated by the interesting morphology exhibited by  MSXG52 in the near- and mid-infrared bands, we carried out molecular lines, and radio continuum observations in order to unveil the nature of this source in the context of the currently accepted massive star formation evolutionary models. In this paper, we report the characterization of the molecular and the ionized gas related to  MSXG52, based on radio continuum observations at 6~GHz carried out with Karl Jansky Very Large Array (JVLA, USA) and on molecular line data obtained using Atacama Submillimeter Telescope Experiment (ASTE, Chile). 

\section{Observations and data reduction}

\subsection{Molecular observations}

The molecular line observations were carried out on August 27 and 28,
2015 with the 10m Atacama Submillimeter Telescope Experiment
\citep[ASTE;][]{eza04}. We used the CATS345~GHz band receiver, which
is a side-band separating SIS receiver remotely tunable in the LO frequency
range of 324-372~GHz. We simultaneously observed $^{12}$CO J=3$-$2 at
345.796~GHz and HCO$^+$ J=4$-$3 at 356.734~GHz, mapping a region of
2$^{\prime}\times 2^{\prime}$~ centered at the position of MSXG52.
 We also observed $^{13}$CO J=3$-$2 at 330.588~GHz and CS J=7$-$6 at
342.883~GHz towards a region of 1\farcm5 $\times$ 1\farcm5 with
the same center. The mapping grid spacing was 10$^{\prime\prime}$~in
both cases, and the integration time was 55~sec ($^{12}$CO and
HCO$^+$) and 95~sec ($^{13}$CO and CS) per pointing. We also performed
a single pointing of C$^{18}$O J=3$-$2 at 329.330~GHz towards the same
center with an integration time of 19~min.  All the observations were
performed in position-switching mode.  We used the XF digital spectrometer
with a bandwidth and spectral resolution set to 128~MHz and 125~kHz,
respectively. The velocity resolution was 0.11~kms$^{-1}$ and the
half-power beam-width (HPBW) was about 22$^{\prime\prime}$~for all
observed molecular lines. The system temperature varied from T$_{\rm
  sys}$ = 150 to 200~K. The main beam efficiency was $\eta_{\rm mb}
\sim$0.65. 

The data were reduced with NEWSTAR\footnote{Reduction software based on AIPS developed at NRAO, ex-
tended to treat single dish data with a graphical user interface
(GUI),} and the spectra
processed using the XSpec software package\footnote{XSpec is a spectral line reduction package for astronomy which
has been developed by Per Bergman at Onsala Space Observatory}.
All the spectra were Hanning-smoothed to improve the
signal-to-noise ratio. The baseline fitting was carried out using
second order polynomials for the $^{12}$CO, $^{13}$CO, and C$^{18}$O
transitions and third-order polynomials for the HCO$^+$
transition. The resulting rms noise of the observations was
about 0.15~K for $^{13}$CO J=3$-$2 and CS J=7$-$6, 0.07~K for
$^{12}$CO J=3$-$2 and HCO$^+$ J=4$-$3 and 0.04~K for C$^{18}$O J=3$-$2
transitions.

\subsection{Radio continuum observations}

The radio continuum observations towards MSXG52
were performed in a single pointing with the Karl G. Jansky Very Large
Array (JVLA) in its C configuration, on February 11, 2016
(project ID:16A-058) for a total of 40 minutes on-source integration
time. We used the wide-band 4-8~GHz receiver system centered at 5.5 and
6.5~GHz, which consists in 16 spectral windows with a bandwidth of
128~MHz each, spread into 64 channels. Data processing was carried out
using the CASA and Miriad software packages, following standard
procedures.  The source J1331+305 was used for primary flux density
and bandpass calibration, while phases were calibrated with
J1931-2243. We reconstructed an image
centered at 6~GHz with a band-with of 2~GHz using the task MAXEN in
MIRIAD, which performs a maximum entropy deconvolution algorithm on a
cube. The resulting synthesized beam has a size of $4\farcs 8
\times 4\farcs 1$, and the rms noise of the final map is 40~$\mu$Jy beam$^{-1}$.

\section{Results and Discussion}

\subsection{Molecular gas and dust}
\label{mol}

\begin{figure}
   \centering
   \includegraphics[width=9cm]{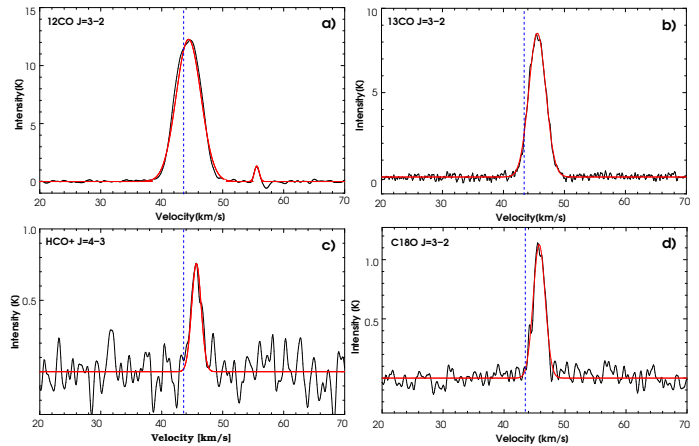}
   \caption{$^{12}$CO J=3$-$2 (a), $^{13}$CO J=3$-$2 (b), HCO$^+$ J=4$-$3 (c), and C$^{18}$O J=3$-$2(d) spectra obtained towards the position of the  MSXG52. The dashed line indicates the radio recombination line related to G52. The Gaussian used to fit the spectra is shown in red.}
              \label{perfiles}
    \end{figure}

\begin{table} 
\caption{Parameters derived from a Gaussian fitting of the spectra shown in Fig. \ref{perfiles}.} 
\centering
\begin{spacing}{1.5}
\begin{tabular}{cccc}
\hline 
\hline
Transition  &  T$_{mb}$ [K] & V$_{LSR}$ [kms$^{-1}$] & $\Delta$v [kms$^{-1}$]\\
\hline 
\hline
$^{12}$CO J=3--2  & 12.3 $\pm$ 0.6 & 44.6 $\pm$ 0.8 & 4.5 $\pm$ 0.6\\ 
                        & 1.3 $\pm$ 0.2 & 55.6 $\pm$ 0.4 & 0.9 $\pm$ 0.4\\ 
              
\hline 
$^{13}$CO J=3--2  & 8.1 $\pm$ 0.5 &  45.4 $\pm$ 0.5 & 3.4 $\pm$ 0.4\\ 
                                       
\hline
C$^{18}$O J=3--2 &  1.2 $\pm$ 0.3 & 45.7 $\pm$ 0.6 & 3.1 $\pm$ 0.5\\
\hline
HCO$^+$ J=4--3  &  0.7 $\pm$ 0.2 & 45.5 $\pm$ 0.4 & 2.3 $\pm$ 0.6\\
\hline
\label{spectra}
\end{tabular}
\end{spacing}
\end{table}

Figure \ref{integrados} shows the averaged velocity maps of $^{12}$CO J=3$-$2, $^{13}$CO J=3$-$2, and HCO$^+$ J=4$-$3 emissions between 42 and 48~kms$^{-1}$. The  $^{12}$CO J=3$-$2 and $^{13}$CO J=3$-$2 emission distributions perfectly match the pillar structure with a conspicuous molecular clump placed towards the head of this structure. MSXG52 (indicated as a green cross in the figure) appears placed onto the border of this clump that faces the \ion{H}{II} region. The HCO$^+$ J=4$-$3 emission appears concentrated  towards the head of the pillar in coincidence with the position of the Red MSX source. Figure \ref{perfiles} shows the $^{12}$CO J=3$-$2, $^{13}$CO J=3$-$2, HCO$^+$ J=4$-$3, and C$^{18}$O J=3$-$2 spectra taken towards the position of  MSXG52 and Table \ref{spectra} shows the derived parameters from Gaussian fits. 

Column density towards the position of the MSXG52 was derived from the $^{13}$CO and C$^{18}$O data by assuming a filled beam and a uniform excitation temperature, common to both tracers (local thermodynamic equilibrium  assumption), within the beam. We first derive the $^{13}$CO opacity, $\tau_{13}$, based on the following equation:

\begin{equation}
\frac{\rm T_{mb}(^{13}CO)}{\rm T_{mb}(C^{18}O)}=\frac{1-{\rm exp}(-\tau_{13})}{1-{\rm exp}(-\tau_{18})}
\end{equation} 

\noindent where we consider T$_{\rm mb}$ at the position of MSXG52. Assuming $\tau_{13}=7.4\tau_{18}$, based on the abundance ratio [$^{13}$C][$^{16}$O]/[$^{12}$C][$^{18}$O]=7.4 estimated at a galacto-centric radius of $D_{GC} = 5.5$~kpc \citep{wil94}, we derive a $\tau_{13} \sim 0.2$, and a $\tau_{18} \sim 0.02$, which shows that both transitions are optically thin towards the position of the MSX source. The excitation temperature can then be derived from the equation of radiative transfer applied to the $^{13}$CO J=3$-$2 transition,

\begin{equation}
T_{mb}(^{13}{\rm CO})=\frac{h\nu}{k}\left(\frac{1}{{\rm exp}(\frac{h\nu}{kT_{ex}})-1}-\frac{1}{{\rm exp}(\frac{h\nu}{kT_{BG}})-1}\right) \times (1-{\rm exp}(-\tau_{13})),
\end{equation}

\noindent where h$\nu/k$=15.87~K and $T_{BG}$=2.7~K. We obtain a $T_{ex}$ of about 52~K.
Given $T_{ex}$ and  $\tau_{13}$, the column density of the $^{13}$CO can be derived from 
\citep[e.g.][]{buc10},

\begin{equation}
{\rm N}(^{13}{\rm CO})=8.28 \times 10^{13}
e^{\frac{h\nu}{k T_{ex}}}\frac{T_{ex}+0.88}{1-exp(\frac{-h \nu}{k T_{ex}})}\int{\tau_{13} {\rm dv}}
\label{N13CO}
\end{equation}

\noindent Taking into account that $^{13}$CO J=3$-$2 transition is  optically thin, the following approximation can be used,

\begin{equation}
\int{\tau {\rm dv}}=\frac{1}{J(T_{ex})-J(T_{BG})}\int{\rm{T_{mb}} {\rm dv}}
\end{equation}

\noindent with

\begin{equation}
J(T) = \frac{h\nu/k}{exp(\frac{h\nu}{kT})-1}.
\end{equation}

\noindent where $\int{\rm{T_{mb}} {\rm dv}} \sim$ 28~K kms$^{-1}$. From the estimated N($^{13}$CO) $\sim 2.1 \times 10^{15}$~cm$^{-2}$, and assuming the [H$_2$]/[$^{13}$CO] ratio of $77 \times 10^4$  \citep{wil94}, we derive a H$_2$ column density of $\sim 1.6 \times 10^{21}$~cm$^{-2}$.

In order to achieve a characterization of the ambient conditions that includes the more external layers of gas  towards this region, we derive the excitation temperature and the column density using  the $^{12}$CO J=3$-$2 and $^{13}$CO J=3$-$2 transitions. Based on the ratio  T$_{\rm mb}$($^{12}$CO)/T$_{\rm mb}$($^{13}$CO) of about 1.5, we estimate opacities of about 55 and 1 for the $^{12}$CO J=3$-$2 and $^{13}$CO J=3$-$2 lines, respectively, and derive a T$_{ex} \sim 20$~K. Using equation \ref{N13CO} and assuming a canonical [$^{12}$CO]/[$^{13}$CO] isotope abundance ratio of 50, we derive a N(H$_2$) of about $5 \times 10^{21}$~cm$^{-2}$. Thus, comparing these results with those obtained for $^{13}$CO J=3$-$2 and C$^{18}$O J=3$-$2 transitions, under LTE conditions (T$_{\rm ex} \sim$ T$_{\rm kin}$), we suggest kinetic temperature and H$_2$ column density gradients across the molecular clump, with the hotter and less dense gas in the inner region. This scenario is in agreement with the presence of a proto-star at its latest stages of evolution embedded in the molecular clump.

By the other hand, from the far-infrared and submillimeter continuum emission of the dust, we estimate a  temperature of the region that can be compared with those obtained above. Under the assumption that the dust radiates as a gray-body characterized by a single temperature $T_d$ and that the emission is optically thin, the flux at a frequency $\nu$ is  $S_\nu \propto N_d k_\nu B_\nu(T_d) $ \citep{and12}, where $B_\nu(T_d)=2 h \nu^3 / c^2 [exp(h \nu / (k_B T_d))-1]^{-1}$ is the Planck function, $k_\nu=k_{\nu 0}(\nu / \nu_{0})^{\beta}$ is the dust opacity (with $\beta=2$, as observed in a large sample of galactic \ion{H}{II} regions) and $N_d$ is the dust column density. The above equation can be written in terms of the H$_2$ column density as

\begin{equation}
S_{\nu} = \mu\, m_H \, N({\rm H_2}) \, k_{\nu 0}(\nu / \nu_{0})^{\beta} \, B_{\nu}(T_d).
\label{eqSED}
\end{equation}

\begin{figure}[ht]
\centering
\includegraphics[width=10cm]{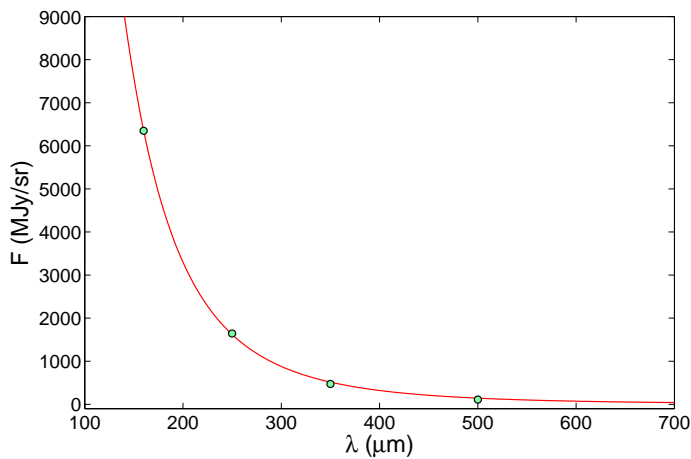}
\caption{Far-infrared and submillimeter continuum fluxes of MSXG52 in the {\it Herschel} 160, 250, 350 and 500 $\mu$m bands. The
red line is the best-fit gray-body model (Eq. \ref{eqSED}) corresponding to a dust temperature of $\sim48$~K and H$_2$ column
density of $\sim2\times10^{21}$~cm$^{-2}$. }
\label{figSED}
\end{figure}

\noindent For a frequency $\nu_{0}=1$~THz and a gas-to-dust ratio of 100, we take $k_{\nu 0}=0.1$ cm$^2$ g$^{-1}$. We adopt a mean molecular weight $\mu=2.8$, corresponding to a relative helium abundance of 10$\%$. We use {\it Herschel} observations to derive the fluxes at different bands and fit the SED of the dust emission with Eq. \ref{eqSED}. Observations \#1342231341 and \#1342231342 cover the field around 
G52 and we used level 2.5 data, which combine both observations in a single set of images.  We fitted the SED using {\it Herschel}-PACS 160 $\mu$m and {\it Herschel}-SPIRES 250, 350 and 500 $\mu$m bands. 
We estimated the flux of the more intense pixel at the four {\it Herschel} images, which corresponds to the position of MSXG52, and subtracted the contribution of the background emission by measuring the mean flux from a circular region centered at $l\sim52.95^\circ$, $b\sim-0.48^\circ$ with a radius of $35^{\prime\prime}$. The background-subtracted peak fluxes are: 6349, 1644, 473 and 111 MJy~sr$^{-1}$ at 160, 250, 350 and 500 $\mu$m, respectively. Flux uncertainties of the {\it Herschel}-PACS and -SPIRES photometers are estimated in $\sim5\%$ (\citealt{bal14} and \citealt{bendo13}, respectively). 
We fitted the data using Eq. \ref{eqSED} leaving $N({\rm H_2})$ and $T_d$ as free parameters. 
The best-fit dust temperature was $T_d=48.1\pm9.7$~K and H$_2$ column density 
$N{\rm(H_2)}=(2.2\pm1.0)\times10^{21}$~cm$^{-2}$, with quoted errors corresponding to the 95\% confidence range. These results support the estimations derived from the molecular observations towards the gas in the inner region. In Fig. \ref{figSED} we plot the fluxes and the best-fit gray-body model.  

\subsubsection{Molecular outflows}
\label{outflow}

\begin{figure}
   \centering
   \includegraphics[width=9cm]{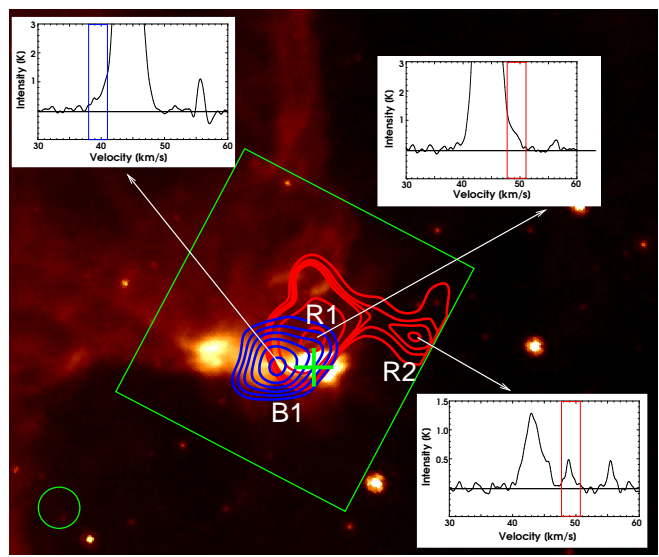}
   \caption{{\it Spitzer}-IRAC image at 8~$\mu$m of MSXG52. The blue and red contours represent the $^{12}$CO J=3$-$2 emission averaged from 36 to 42 kms$^{-1}$ (blue
lobe), and from 48 to 51 kms$^{-1}$ (red lobe), respectively. The blue contours are at 0.10, 0.12, 0.14, 0.16, and 0.18~K and the red ones are at 0.10, 0.12, 0.13, 0.14, 0.20, 0.30~K. We show the spectra towards the center of features B1, R1, and R2. The green cross indicates the location of MSXG52. The green box shows the area mapped with ASTE at this transition. The green circle represent the beam of the molecular observations.}
     \label{molec_out}
    \end{figure}

MSXG52 exhibits in the 8~$\mu$m and $K_s-$band images a morphology very suggestive of outflow activity in the region. A detailed inspection of the whole $^{12}$CO J=3$-$2 data cube revealed the presence of spectral wings in some $^{12}$CO J=3$-$2 spectra (shown in Figure \ref{molec_out}), likely associated with molecular outflows. Figure \ref{molec_out} shows {\it Spitzer}-IRAC image at 8~$\mu$m of MSXG52. The blue and red contours represent the $^{12}$CO J=3$-$2 emission averaged from 36 to 42 kms$^{-1}$ (blue lobe), and from 48 to 51 kms$^{-1}$ (red lobe). The feature B1 related to the blue-shifted emission appears located in projection onto the jet-like structure as seen at 8~$\mu$m. The spectrum obtained towards the center of this lobe exhibits a blue wing highlighted with a blue rectangle. The red-shifted emission exhibits a more complex morphology with two features R1, and R2 that are misaligned with respect to the axis of the jet-like structure.  In particular, the spectrum obtained towards the center of source R2 exhibits a velocity component centered at 49~kms$^{-1}$, which could be due to a red-shifted bullet of molecular gas.
The velocity component centered at 55~kms$^{-1}$, which is observed in the three spectra, corresponds to molecular gas associated with the pillar structure.

To roughly estimate the outflow mass, following \citet{ber93}, we calculate the H$_2$ column density from

\begin{equation}
{\rm N(H_2)}=2.0 \times 10^{20}\frac{W(^{12}{\rm CO})}{{\rm K~kms}^{-1}}({\rm cm}^{-2})
\label{nh2_12co}
\end{equation}

\noindent where {\it W}($^{12}$CO) is the $^{12}$CO J=3$-$2 integrated intensity along the intervals mentioned above. Then, the mass was derived from

\begin{equation}
{\rm M}=\mu~m_H~{\rm D}^2 \Omega \sum_i {\rm N_i(H_2)}
\end{equation}

\noindent where $\Omega$ is the solid angle subtended by the beam size, m$_H$ is the hydrogen mass, $\mu$=2.8 corresponds to a relative helium abundance of 10\%, and D is the distance. We summed over all beam positions belonging to the lobes, which yields the mass for the red- and blue-shifted outflows: M$_{red} \sim 6$~M$_{\odot}$ and M$_{blue} \sim 5$~M$_{\odot}$, in agreement with typical values for  massive molecular outflows \citep{wu04}.

From a simple inspection of the calibrated near-IR spectrum in the H$+$K bands towards MSXG52 \citep{coop13}, kindly provided by Lumsden S., it can be appreciated a prominent Br$\gamma$ emission line, and no evidence of H$_{2}$ 1--0 S(1) and [FeII] lines. This Br$\gamma$ emission might arise from ionized stellar wind generated in an evolved massive proto-star \citep{Bik06,kraus08} or from a compact \ion{H}{II} region in photoionization equilibrium \citep{kro81}. Additionally, the non--detection of H$_{2}$ 1--0 S(1) and [FeII] lines would suggest the absence of collimated jets shocking the inner regions of a YSO envelopment \citep{reip00,bally07}. 
Thus, the lack of evidence of collimated jets would suggest that the outflow activity in MSXG52 would be related to a wide-angle ionized stellar wind.  \citet{wei06} suggest that a wide-angle stellar wind still plays an important role in driving molecular outflows.

\begin{figure*}
\centering
    \includegraphics[width=14cm]{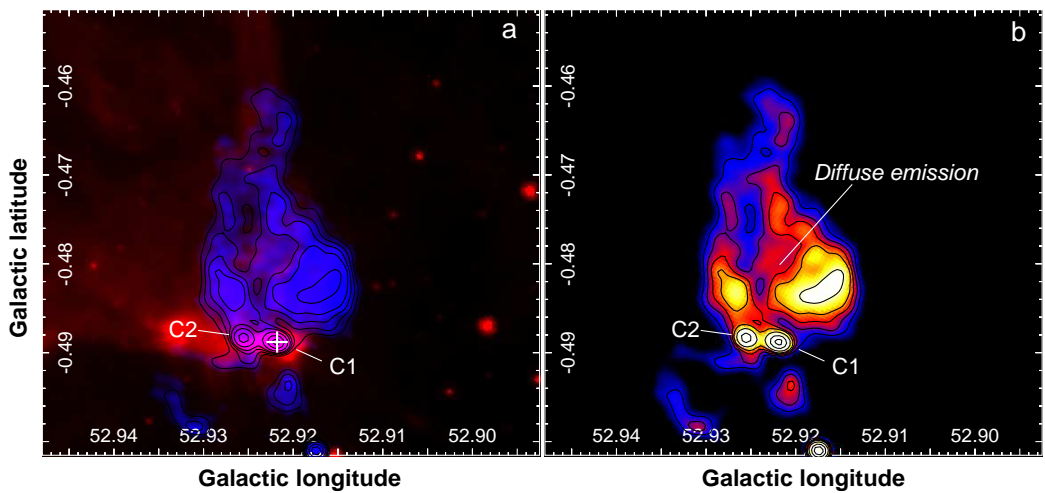}
   \caption{(a) Two-color image with 8~$\mu$m in red and 6~GHz in blue.  The cross indicates the position of  MSXG52. (b) Radio continuum emission at 6~GHz. The scale goes from 0.08 to 0.3~mJy~beam$^{-1}$. Levels are at 0.12, 0.14, 0.16, 0.19, 0.25, and 0.35~mJy~beam$^{-1}$.  C1 and C2 show the position of the two compact radio sources mentioned in the text.} 
              \label{jvla}
    \end{figure*}

\subsection{The ionized gas}
\label{radio}

\begin{figure}
   
   \includegraphics[width=9cm]{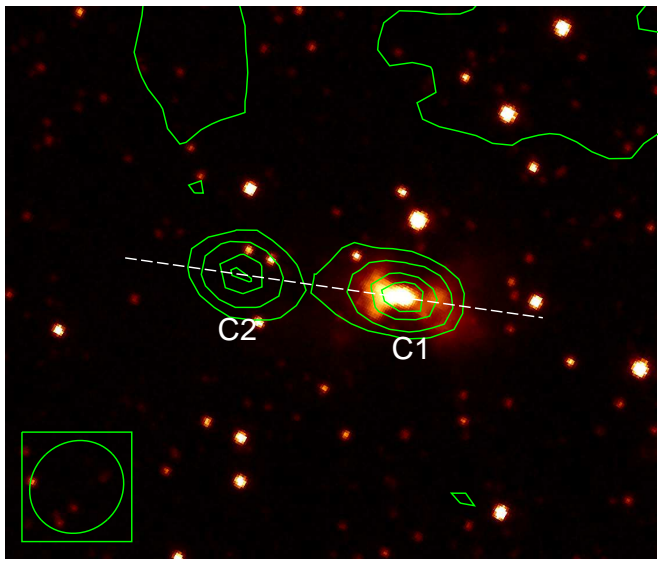}
   \caption{Near-infrared emission at the K$_s-$band as extracted from the UKIDSS Survey. The green contours represent the radio continuum emission at 6~GHz at 0.2, 0.3, 0.4, and 0.5~mJy~beam$^{-1}$. The beam of the radio continuum observation is shown in the bottom-left corner.}
              \label{jvla+K}
    \end{figure}

Figure \ref{jvla}-a shows a two-color image with the 8~$\mu$m emission in red and the radio continuum emission at 6~GHz in blue. It can be noticed that the diffuse radio continuum emission perfectly matches the north border of the pillar-like structure as seen at 8~$\mu$m. This  emission seems to arise from the ionized gas related to the large \ion{H}{II} region that has been stalled against the pillar, illuminating it. This is in agreement with the assumption that the exciting star(s) of the large \ion{H}{II} region is(are) located in the bulk of emission at 24~$\mu$m (see Fig. \ref{intro}). By the other hand, there are two conspicuous compact radio sources lying along the southern border of the pillar. One of them, named C1, coincides in position with MSXG52. The other source, C2, is seen in projection onto the tail of the jet-like feature. As can be appreciated from Fig. \ref{jvla+K}, source C1 shows a slightly elongated morphology at 6~GHz with respect to source C2. In particular, C1 embraces perfectly the  UKIDSS source UGPS J193054.62$+$172842.0, which is the near-infrared counterpart of MSXG52 at the K$_s-$band. At this band, it can be noticed the young star at the center of the hourglass-like nebula, which suggests the presence of cavities cleared in the circumstellar material. Moreover, the young star coincides in position with the peak of the radio emission of C1. The axis of symmetry connecting  both compact radio sources (white dashed line in Fig. \ref{jvla+K}) shows a perfect alignment with the axis of symmetry of the hourglass-like nebula and with the jet-like feature seen at 8~$\mu$m, which suggests that these structures are related.

The primary mechanism of the radio continuum emission for MYSOs is thermal free-free emission from ionized gas. This may either be in the form of  thermal jets, ionized stellar winds, and/or photo-ionized compact \ion{H}{II} regions \citep[e.g.][]{gib07, rod12}. In turn, these compact photo-ionized regions can be
divided in hyper- and ultra-compact \ion{H}{II} regions, which at 6~GHz are in the optically thick and optically thin regime, respectively \citep{kur05}.
The estimated radio flux density for C1 at 6~GHz of about 0.9~mJy, is in agreement with all these scenarios \citep{rod12}.

In this context, a first clue to the origin of the ionized gas associated with MSXG52 and therefore of the nature of the source, could be provided by a comparison between the Br$\gamma$ intensity and the radio continuum emission at 6~GHz. From the NIR spectrum kindly provided by Lumsden S. \citep{coop13}, we derived a Br$\gamma$ intensity of about $0.13 \times 10^{-12}$ ergs cm$^{-2}$ s$^{-1}$. Assuming a ratio Br$\alpha$/Br$\gamma$ of about 2.33 \citep{wyn84}, and a radio continuum emission at the optically thin regime, we estimated the ratio Br$\alpha$/S(5~GHz) of about $0.033 \times 10^{-12}$ ergs cm$^{-2}$ s$^{-1}$ mJy$^{-1}$, which is in agreement with the predicted value of $0.026 \times 10^{-12}$ ergs cm$^{-2}$ s$^{-1}$ mJy$^{-1}$ for an optically thin \ion{H}{II} region in photoionization equilibrium \citep{sne86}. This result would rule out the presence of an optically thick envelope surrounding a star arising in a ionized stellar wind \citep{kro81}. It is important to mention that considering the radio continuum emission at the optically thick regime \citep{kur05}, only introduces a factor of about 1.5 in the Br$\alpha$/S(5~GHz) ratio derived above. 
This result suggests that the radio source C1 would be the ionized gas of an optically thin  UC \ion{H}{II} region associated with a young massive star. If this is the case, we can conjecture upon the spectral type of a probable single ZAMS star based on the estimated radio flux density for C1. The number of photons needed to keep an \ion{H}{II} region ionized, in an optically thin regime, is given by ${\rm N_{uv}=0.76 \times 10^ {47} T_4^{-0.45} \nu_{GHZ}^{0.1} S_{\nu} D_{kpc}^2}$ \citep{cha76}, where T$_4$ is the electron temperature in units of $10^4$~K, D$_{\rm kpc}$ is the distance in kilo-parsecs, ${\rm \nu_{GHz}}$ is the frequency in GHz, and S$_{\rm \nu}$ is the measured total flux density in Jansky.
We assumed an electron temperature of T=10$^4$~K and a distance of 7.2~kpc. We derived a total amount of ionized photons in C1 of about ${\rm N_{uv}=(4.2 \pm 2.2)\times 10^{45} ph s^{-1}}$. Based on \citet{ave79} and \citet{mar05}, we conclude that the spectral type of the exciting star of this wimpy UC \ion{H}{II} region should be B0.5V.

By the other hand, the presence of molecular outflows related to MSXG52, together with the slightly elongated morphology of the associated radio continuum emission (source C1), would suggest that the source is transiting an earlier evolutionary stage. If this is the case, given the lack of evidence of collimated jets in the near-infrared spectrum (see Section \ref{outflow}), we could find an explanation to its elongated morphology in the work of \citet{ket07}, which suggests that at some point of its evolution, a HC \ion{H}{II} region exhibits a bipolar morphology that accounts for the presence of wide-angle ionized stellar winds flowing through the poles.  The hourglass morphology of the source at the K$_s-$band is in agreement with this scenario. Thus, in the context of the evolutionary models for high-mass stars proposed by \citet{beu05}, we suggest that MSXG52 could be at the evolutionary stage in which the jets are giving way to the wide-angle ionized winds together with a photo-ionized  region around the young star.  Thus, the jet-like feature seen at 8~$\mu$m  and the hourglass nebula detected  at the K$_s-$band would be generated by the action of the UV photons and winds that escape from the polar regions of  the young star.

Regarding the nature of the source C2, its perfect alignment with the axis of symmetry of the radio source C1, together with the absence of an IR counterpart, would suggest that this radio source is tracing an ionized stellar wind breaking out the central source in an episodic ejection event. However, the possibility that C2 is related to a companion MYSO could not be ruled out. 

\section{Summary and Conclusion}

We study the molecular and ionized gas in the Red MSX source G052.9221$-$00.4892. The $^{12}$CO J=3$-$2 emission shows the presence of misaligned high-mass molecular outflows. In particular, the blue lobe coincides in position with a jet-like structure, detected at 8~$\mu$m, that arises from the central source. However, from the near-IR spectrum, we did not find signatures of collimated jets related to this source. The radio continuum emission at 6~GHz shows a slightly elongated radio compact source  in positional coincidence with the Red MSX source. This radio source perfectly matches the hourglass nebula seen at the K$_s-$band. 
Thus, the presence of molecular outflows related to the Red MSX source G052.9221$-$00.4892, the slightly elongated morphology of the associated radio continuum emission matching the hourglass-like morphology exhibited by the source at the K$_s-$band, and the lack of evidence of collimated jets, suggest that the source could be transiting the earlier evolutionary stage of HC \ion{H}{II} region with the presence of a bipolar wide-open angle ionized wind. \citet{ket07} and \citet{beu05} predict an evolutionary stage in the massive stellar formation in which the star reaches the ZAMS and continues with the accretion mainly through the equatorial plane causing an outwards flow of ionized gas and winds emanating from the poles. Our studied source could be an observational evidence of this stage. Taking into account that the cataloged HC \ion{H}{II} regions are scarce, finding and characterizing  sources like the studied in this work is very important to advance in the understanding of the massive star formation and how the stars build up their masses once they have reached the ZAMS.

By the other hand, according to a comparison between the Br$\gamma$ intensity and the radio flux density at 6~GHz, the source would be in a more evolved evolutionary stage of an optically thin UC \ion{H}{II} region in photoionization equilibrium. If this is the case, from the estimated radio flux density, we can conjecture upon the spectral type of its exciting star which would be a B0.5V.

\begin{acknowledgements}

We acknowledge the anonymous referee for her/his helpful comments and suggestions. We wish to thank to S. Lumsden for kindly provide us with the near-infrared spectrum of the Red MSX source G052.9221$-$00.4892.
The ASTE project is led by Nobeyama Radio Observatory (NRO), a branch of National Astronomical Observatory of Japan (NAOJ), in collaboration with University of Chile, and Japanese institutes including University of Tokyo, Nagoya University, Osaka Prefecture University, Ibaraki University, Hokkaido University, and the Joetsu University of Education.  M.O., S.P., E.G., and A.P. are members of the Carrera del investigador cient\'ifico of CONICET, Argentina. M.C.P. is a doctoral fellow of CONICET, Argentina. This work was partially supported by grants awarded by CONICET, ANPCYT and UBA (UBACyT) from Argentina. A.P. acknowledges the support from the Varsavsky Foundation. M.R. wishes to acknowledge support from FONDECYT(CHILE) grant N$^{\rm o}$1140839.
S.P. and A.P. are grateful to Dr. Takeshi Okuda for the support received during the ASTE observations.

\end{acknowledgements}

\bibliographystyle{aa} 
\bibliography{biblio} 

\end{document}